\documentclass[aps,prl,twocolumn,groupedaddress]{revtex4}
\usepackage{graphicx}

\begin{document}
\newcommand{\bfr}{{\bf r}}
\newcommand{\bft}{{\bf t}}
\newcommand{\bfv}{{\bf v}}
\newcommand{\Sp}{{\rm Sp}}
\newcommand{\bffv}{{\bf f}_{\rm v}}
\newcommand{\bffe}{{\bf f}_{\rm e}}
\newcommand{\SFe}{F_{\rm e}}

\title{Rotational dynamics of a soft filament: wrapping transition and
propulsive forces}

\author{N. Coq, O. du Roure, J. Marthelot, D. Bartolo \& M. Fermigier }
\affiliation{Laboratoire de Physique et M\'{e}canique des Milieux
H\'{e}t\'{e}rog\`{e}nes,ESPCI-CNRS UMR7636-Paris 6-Paris 7, 10 rue
Vauquelin, 75005 Paris,
France}

\date{\today}

\begin{abstract}
We analyze experimentally the shape of a long elastic filament rotating in
a viscous liquid. We identify a continuous but sharp transition from a
straight to  an helical shape, resulting from the competition between
viscous stresses and elastic forces. This induced helicity generates a
propulsive force along the axis of rotation. In addition, we show that the
shape transition is associated with an unstable branch in the force-torque
relation, confirming the numerical predictions of Manghi \textit{et
al.}~\cite{netz06}.  A linearized model of the fluid-structure interaction
is proposed to account for all the features of the non-linear filament
dynamics.
\end{abstract}


\maketitle

Many cells use the beating of elastic filaments to swim or to pump
fluids~\cite{bray04}. Prominent examples are the swimming of sperm cells
which propel themselves by exciting propagative deformations along a
single flagellum~\cite{camalet99}, and the pumping of liquid by the
helical motion  of cilia on embryo nodal cells~\cite{takeda06}. Since the
pioneering work of Taylor in the early 50s, the observation of these
fascinating biological machines has inspired numerous studies on the
fluid-structure interaction of flexible filaments with viscous flows.
Moreover, recent advances in the construction of complex colloidal
assembly~\cite{pine,bibette} and in the coupling of biological machines to
artificial microstructures~\cite{GMWscience07} should allow man-made
swimmers to catch up with microorganisms. A promising example has already
been  proposed by Dreyfus {\it et.al.} who have quantitavely studied the
propulsion of the first artificial flexible
micro-swimmer~\cite{dreyfus04}. So far, special attention has been paid to
the thrust  produced by the periodic and planar oscillations of an
isolated flagella~\cite{taylor51,lauga07,powers,riveline,pecko,dreyfus04}.
However, in the last two years, a set of numerical and theoretical
works~\cite{netz06,julicher06} has been devoted to another propulsion
mechanism, the rotation of a single tilted flexible rod.

In this paper we present an experimental realization of this system. We
show that, increasing the angular velocity, $\omega$, the filament
undergoes a sharp but continuous shape transition from a linear to an
helical shape tightly wrapped around the rotation axis. We show that this
collapse of the flexible rod is solely ruled by the interplay between the
elastic forces and the viscous drag acting on it. The relation between the
filament shape, the axial force and the rotation torque acting on the
filament is investigated using a high resolution imaging method and
described quantitatively thanks to a simple model of the fluid-structure
interaction. We also give experimental evidence that a torque-controlled
rotation should lead to strongly non linear and unstable filament
dynamics~\cite{netz06}.
\begin{figure}[htbp]
\begin{center}
\includegraphics[width=0.9\columnwidth]{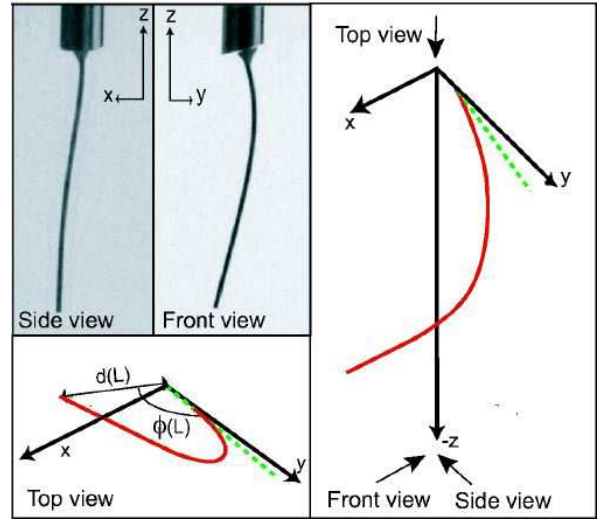}
\caption{Top left: front and side view of a rotating filament. Right:
reconstructed 3D shape of the filament (solid line) and slope at the
anchoring point (dashed line). Bottom left: projection of the filament
shape in the $(x,y)$ plane perpendicular to the rotation axis (solid line)
and slope at the anchoring point (dashed line).}\label{fig1}
\end{center}
\end{figure}

We rotate a flexible filament immersed in a transparent plexiglas tank
(dimensions 20 x 20 x 20 cm$^3$) filled with pure glycerin. The shear
viscosity, $\eta$, of the glycerin has been systematically measured prior
to each experiment. We did not measure any change due to possible
temperature or hygroscopic variations: $\eta=1\,$Pa$.$s.  The filaments
are made of a low modulus polyvinylsiloxane elastomer. Glass capillary
tubes are filled with a mixture of polymer and curing agent containing
dispersed iron carbonyl particles intended to match the density of
glycerin. Once the polymer is cured the glass capillary is broken to
recover a cylindrical elastic rod of radius $a=435\,\mu$m, which length
varies from 2 cm to 10 cm. The Young's modulus $E=0.7$ MPA of each rod was
measured by dynamical mechanical analysis. The filaments are then attached to the axis of an
electric motor delivering a discrete set of rotation speeds ranging from
0.01 to 10 rpm through a gear box. In all our experiments, the motor axis
and the filament at rest make an angle $\theta$ of $15^\circ$.  We
simultaneously take pictures of the rotating filaments in two
perpendicular directions with a 6 MPixels digital camera (Nikon D70).
Eventually, we use a correlation algorithm to detect the coordinates of
the two corresponding projected profiles, which allows for the
determination of the full 3D shape of the distorted rods with a
submillimeter accuracy, Figure~\ref{fig1}.
\begin{figure}
\begin{center}
\includegraphics[width=5cm]{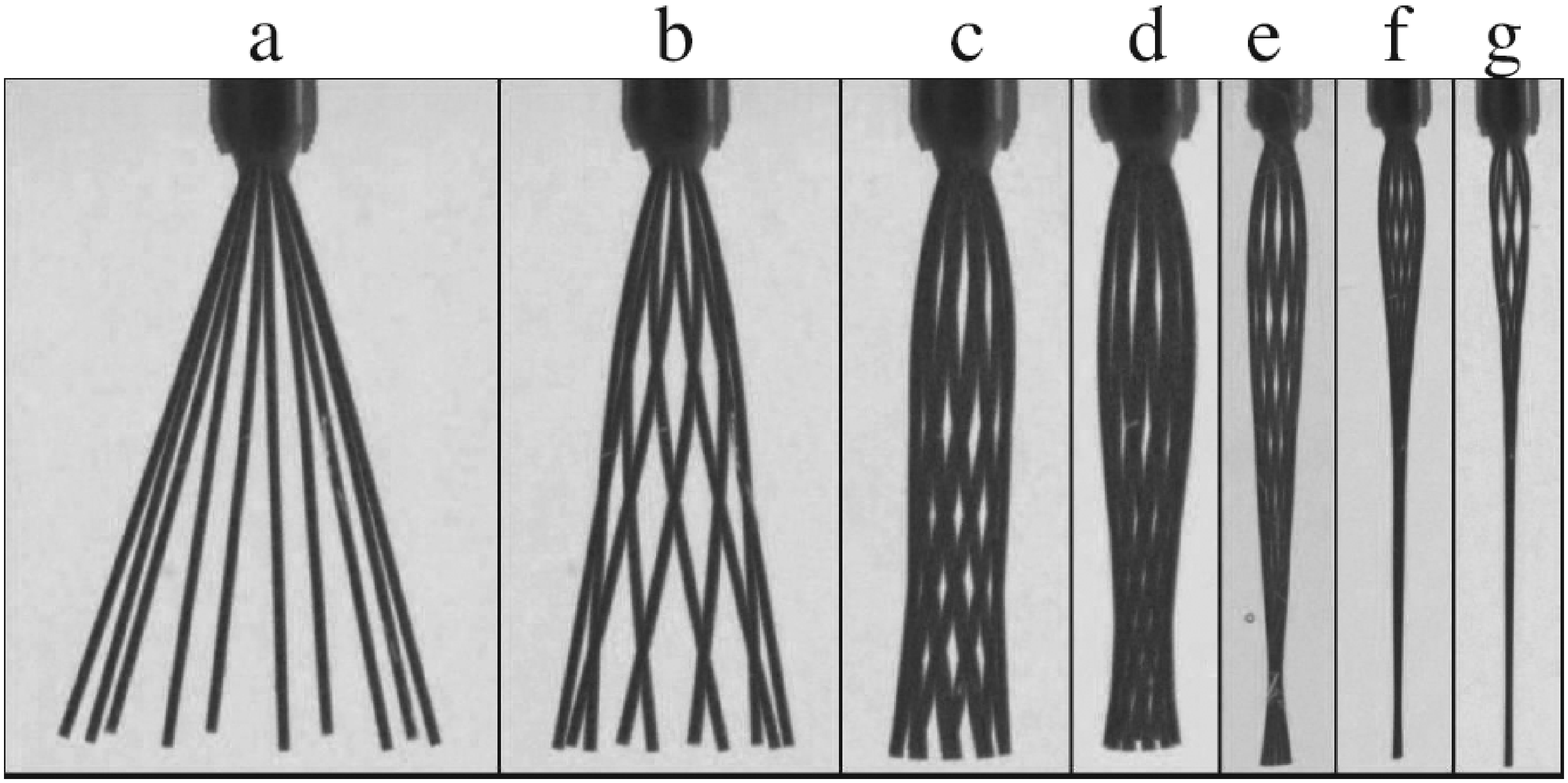}
\includegraphics[width=\columnwidth]{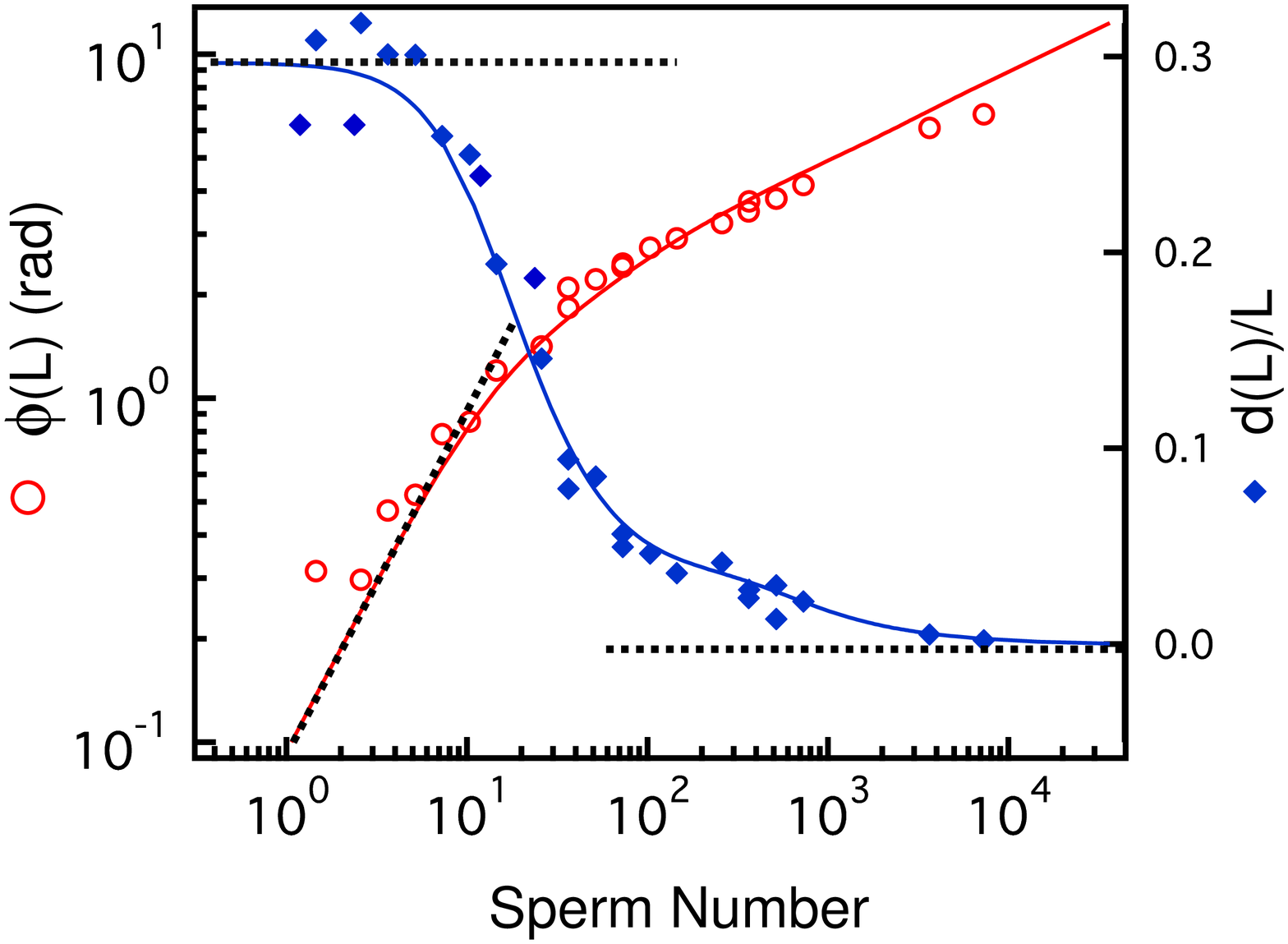}
\caption{Polar angle in the $(x,y)$ plane (circles) and distance to the
axis (diamonds) of the free end of the filament as a function of the
dimensionless angular velocity Sp. Solid lines are solutions of the
linearized deformation equation. Dashed lines: analytical solution in the
low  and high Sp limits, see main text. On top, corresponding shapes of
the rotating filament (superimposed pictures at different times within a
period).}
\label{fig2}
\end{center}
\end{figure}
After a transient regime the rotating filament reaches a stationary shape
and undergoes a rigid body rotation. The shortest rods are hardly deformed
by the viscous flow even when the rotation speed is increased by 3 orders
of magnitude. They rapidly adopt a slightly chiral shape close to the
initial straight and tilted  conformation. The rods with intermediate
lengths display a continuous but sharp transition from an almost straight
to an helical shape when increasing the angular velocity, figure ~\ref{fig2}. The
longest rods are significantly bent by the viscous drag; after a long
transient regime ($\sim1$ hour), they are tightly wrapped around the
rotation axis even at the slowest rotation speed.
In all that follows, we restrain our attention only to the final
stationary shapes.

To go beyond the above qualitative observations, the dimensionless distance to the
rotation axis, $d(L)/L$ and the polar angle $\phi(L)$ of the rod end  are plotted in
figure~\ref{fig2} as a function of the non dimensional rotation speed ${\rm Sp}\equiv \omega
\eta_{\perp} L^4/\kappa$. Sp is commonly referred to as the sperm number,
it compares the period of angular rotation to the elasto-viscous
relaxation time, $\tau=\eta_\perp L^4/\kappa$, of the bending mode of
wavelength $L$~\cite{lauga07,wiggins98}, where $\kappa=\pi E a^4/4$ is the
bending modulus of the filament and $\eta_{\perp}$ is the drag coefficient
for normal motion.  First of all, it is worth noticing that all the
experimental data collapse on the same master curve, which implies that
the deformation of the rods results from the competition between viscous
and elastic forces.
\begin{figure}
\begin{center}
\includegraphics[width=\columnwidth]{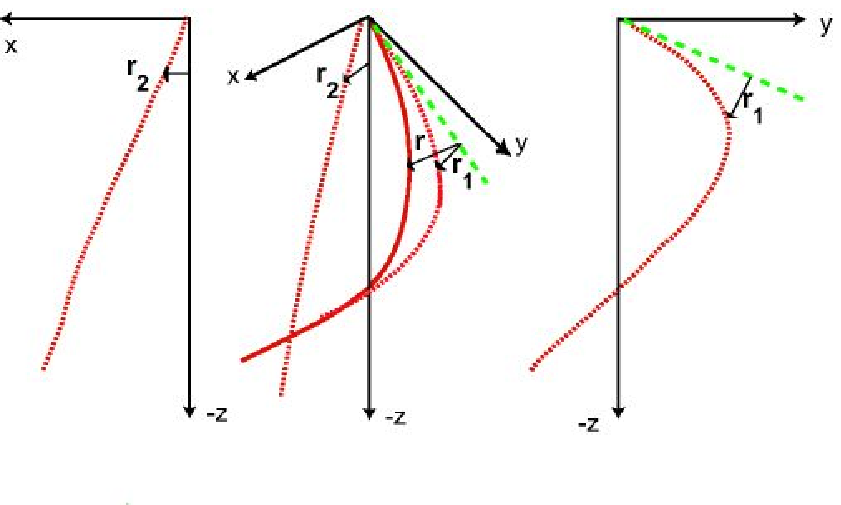}
\caption{Sketch of the filament deformations in the (x,z) and (y,z) planes illustrating the definition of  the displacement field $\bfr (s)=(r_1(s),r_2(s))$. Solid lines:  filament shape in 3D. Dashed lines: undeformed filament. Dotted lines: projections of the filament on the (x,z) and (y,z) planes.}
\label{fig3}
\end{center}
\end{figure}

At low Sp, the polar angle $\phi(L)$ increases linearly with $\omega$
whereas $d(L)/L$ remains constant over more than two decades. Above $Sp
\approx 10$, the variation of the polar angle becomes much weaker.
Conversely, the distance to the rotation axis drops down to a very small
value. Surprisingly, a quantitative description of this wrapping
transition can be performed ignoring both geometrical nonlinearities and
long-range hydrodynamic coupling. To determine the filament shape we
compute the elastic and the viscous forces acting on the flexible rod.
Using a local drag description the viscous force is
\begin{equation}
{\bf f}_{\rm v}=(\eta_\parallel-\eta_\perp)(\bft.\bfv)\bft+\eta_\perp \bfv,\label{force}
\end{equation}
with $\bf t$ the tangent vector, $\eta_\perp=4\pi\eta/\left
[\log(L/a)+\frac{1}{2}\right ]$ and $\eta_\parallel=2\pi\eta/\left
[\log(L/a)-\frac{1}{2}\right ]$ the drag coefficients in the slender body
approximation~\cite{slender}. The elastic force $\bf f_e$ derives from the
bending energy functional written within  the small deformations
approximation ${\cal E}=\frac{1}{2}\int\, \kappa (\partial^2_s
\bfr)^2\,{\rm d}s$, with $s$ the curvilinear coordinate. $\bfr
(s)=(r_1(s),r_2(s))$ is the displacement field normal to the undeformed
filament, see Figure~\ref{fig3}. Ignoring the incompressibility constraint
which would only add extra nonlinear contributions to the linearized
elastic force: $\bffe=-\kappa\partial^4_s\bfr$, the filament shape can
then be exactly computed by solving the force balance equation
$\bffe=-\bffv$ in the frame rotating at $\omega$ around the $z$ axis.
Introducing the penetration length of the bending modes
$l(\omega)\equiv[\kappa/(\eta_\perp\omega\cos\theta)]^{1/4}$ this equation
can be written in the compact form:
\begin{eqnarray}
l^4(\omega)\partial_s^4r_1&=&-r_2-s\tan \theta,\label{Eq1}\\
l^4(\omega)\partial_s^4r_2&=&r_1.\label{Eq2}
\end{eqnarray}
with the torque and force free conditions at $s=L$:
$\partial_s^2\bfr(L)=\partial_s^3\bfr(L)=0$
and the geometrical constraints  on the rotation axis:
$\bfr(0)=\partial_s\bfr(0)=0$.  The excellent agreement between the
theoretical and the measured geometrical parameters plotted in
Figure~\ref{fig2} demonstrates that this simplified approach correctly
captures the main features of the filament dynamics. Although this linear
equation can be solved analytically, the form of the exact solution is so
complex that it is not really insightful. We rather detail here the two
asymptotic regimes $Sp\sim [L/l(\omega)]^4\ll1$ and $Sp\sim
[L/l(\omega)]^4\gg1$ corresponding to almost straight and tightly wrapped
rods respectively. In the limit of large $l(\omega)$ (low speeds), the
solution of the two above equations is:
\begin{eqnarray}
&&r_1=-\frac{L \tan \theta}{120}Sp \left[20\left (s/L\right)^2-(10s/L)^3
+(s/L)^5\right] \; \label{Eq12}\\
&&r_2={\cal O}({\rm Sp}^2) \; \label{Eq22}.
\end{eqnarray}
It then follows that the rotation mostly bends the filament in the flow
direction, the distance $d(L)$ is thus expected to remain constant at low
speed. Conversely, since the filament responds linearly to the viscous
flow, the wrapping angle $\phi\sim r_1(L)/(L\sin\theta)$ increases
linearly with Sp: $\phi=(11/120)Sp$. These two predictions thus correctly
capture the main features of two experimental observations reported in
Figure~\ref{fig2}.
In the limit of small $l(\omega)$ (high speeds), Eqs.~\ref{Eq1} and
\ref{Eq2} reduce to $r_2(s)=-s\tan \theta$ and $r_1=0$. This immediately
tells us that the filament is now completely aligned along the rotation
axis in this high speed regime. More precisely, the flow induces a strong
bending of the filament but the curvature is only localized in a region of
length $l(\omega)$ near the anchoring point on the $z$ axis.  This
explains the surprising collapse seen in our experimental pictures,
Figure~\ref{fig2}.
\begin{figure}
\begin{center}
\includegraphics[width=\columnwidth]{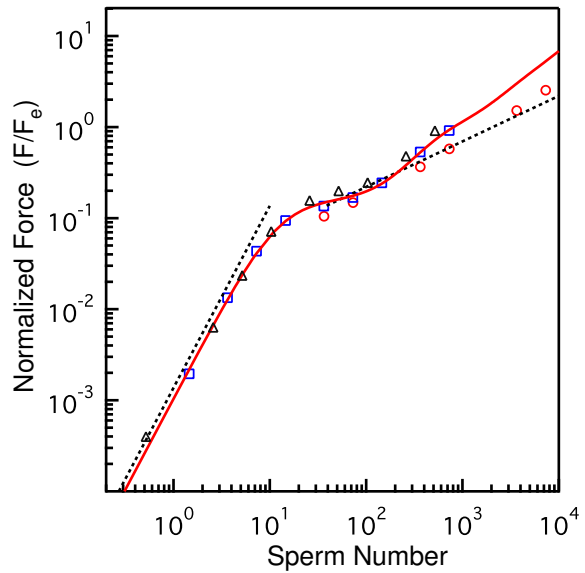}
\caption{Propulsive force normalized by the elastic force as a function of
Sp for three different filament lengths (triangles: $L$= 48 mm, squares:
$L$=52.5 mm, circles: $L$= 96 mm). Dotted lines: Theoretical predictions in the low and high Sp limits given by Eq~\ref{Eq55} and ($F/F_e \sim Sp^{1/2}$) respectively.
Solid line: Force computed from the profiles obtained by solving Eqs.~\ref{Eq1}and ~\ref{Eq2}.}
\label{forcesp}
\end{center}
\end{figure}
Our second main objective is now to assess the impact of this
rotation-induced wrapping  on the (propulsive) axial force $F$ created by
the flow and on the torque required to enforce the stationary rotation.

We first focus our attention on the variation of the axial force in an
angular velocity-controlled experiment.
Our accurate filament detection algorithm enables us to measure axial
force $F=\int {\bf f}_v(s)\cdot {\bf e}_z \,ds$
values as low as $3\,n$N. Contrary to what would be observed with a rigid
filament, we systematically measure a non zero axial force. Moreover the
direction of the force is independent of the sign of the angular velocity.
This can be understood by looking at the symmetry of the deformed
filaments. A positive (resp. negative) $\omega$ induces left- (resp.
right-) handed helical stationary deformations to the initially straight
flexible rods. Besides it is well known that a left- (resp. right-) handed
chiral object rotating in the clockwise (resp. anticlockwise) direction
experiences an upward (resp. downward) viscous force. We can thus
anticipate that the axial force $F$ should increase quadratically with
$\omega$ at least in the low Sp limit. Dimensional analysis then implies
that $F$ should scale as $F_{\rm e}{\rm Sp}^2$, where we define the
elastic force $F_{\rm e}\equiv\kappa/L^2$. To go beyond this scaling
prediction we can compute the total force knowing the filament shape in
the low Sp limit thanks to our simplified linear model (Eq.~\ref{Eq12}):
\begin{equation}
F=\left(1-\frac{\eta_\parallel}{\eta_\perp}\right) \frac{\sin^2\theta\cos \theta F_{\rm e}}{18}{\rm Sp}^2+{\cal O}({\rm
Sp}^3). \label{Eq55}
\end{equation}
This expression is in excellent agreement with our experimental findings
for sperm numbers smaller than $10$, Figure~\ref{forcesp}. This figure
shows that $F$ continuously increases with the dimensionless angular speed
and reveals a second power-law regime in the other asymptotic limit. For
$Sp>10$, the force scales as $F\sim F_e Sp^{1/2}$. We notice that the
crossover between the two power-law behaviors occurs in the narrow range
of Sp where the filament starts bending towards the $z$-axis.

Besides, we have shown that the elastic deformations of the filament are
localized over a length $l(\omega)$ in the tightly wrapped conformations.
Hence, a simple scaling analysis predicts that the axial force experienced
by the filament should scale as $F\sim \eta \omega l^2(\omega)$, or equivalently
$F/F_{\rm e}\sim {\rm Sp}^{1/2}$, which is observed in
Figure~\ref{forcesp}. 

\begin{figure}
\begin{center}
\includegraphics[width=\columnwidth]{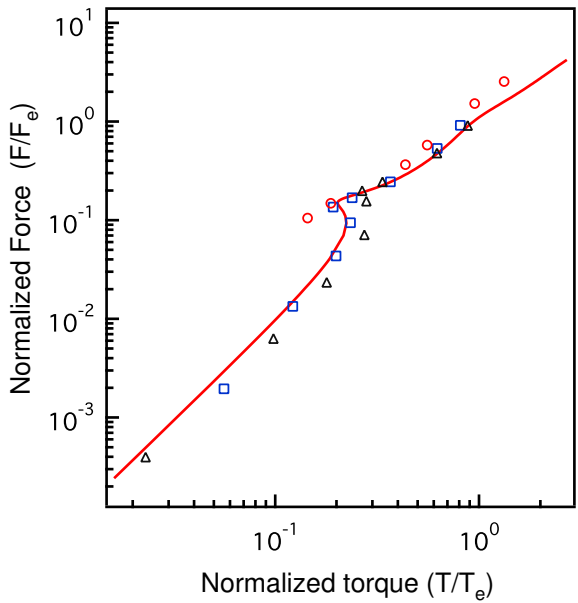}
\caption{Normalized propulsive force vs normalized torque for three different filament lengths (triangles: $L$= 48 mm, squares: $L$=52.5 mm, circles: $L$= 96 mm). Solid line: Force-Torque relation obtained according to our linearized model.}
\label{torque}
\end{center}
\end{figure}
We now come to our last and important results. We have also systematically
computed the viscous torque, $T$, acting on the flexible rods, from the 3D
shape reconstruction: $T=\int\,\left [{\bf f}_{\rm v}\cdot {\bf e}_z
\right ]d(s){\rm d}s$, with $d(s)$ the distance from the $z$ axis. In the
stationary state the measured viscous torque is equal to the torque
delivered by the motor. Hence, we can deduce the evolution of the axial
force in a torque-driven experiment from this measured torque. The axial
force normalized by the elastic force is plotted versus $T/T_{\rm e}$ in
Figure ~\ref{torque}, with the elastic torque $T_{\rm e}\equiv\kappa/L$.
Contrary to what is observed for the force-velocity relation the axial
force is a non monotonic function of the applied torque. We stress on the
surprising decrease of the force with $T$ for $T\sim 0.2 T_{\rm e}$. This
counterintuitive behavior is actually observed for sperm numbers for which
the filament collapses on the $z$ axis. A decreasing branch in the
torque-force diagram implies that a torque-driven filament would undergo a
discontinuous structural transition. This confirms the observations made
by  Manghi {\it et al.} in Stokesians numerical simulations~\cite{MSN06}.
Eventually we also emphasize the remarkable robustness of our simple
modeling in accounting for the fluid-structure interaction. This {\it
linear} model yields again an excellent prediction of these strongly {\it
nonlinear} variations of the force with the driving torque
(Figure~\ref{torque}).

From a design perspective, the self-induced helicity of an elastic
flagellum could be an efficient strategy to drive artificial swimmers. On
the one hand, operating at a constant rotational speed ensures a very stable
swimming speed. On the other hand, choosing a working point close to the
discontinuous shape transition would allow for strong accelerations
triggered by a slight variation of the torque command.  An interesting
issue which goes beyond the scope of this paper deals with the efficiency of
such a propulsive mechanism, both in the pumping and the swimming
regimes \cite{avron}.

\begin{acknowledgments}
H. Stone and R. Netz are gratefuly acknowledged for stimulating
discussions. We thank N. Champagne, E. La\"ik and L. Gani for help with the
experiments.
\end{acknowledgments}

\noindent \\While we were completing this work, we became aware of a very
similar study from K. Breuer's group~\cite{KB07}.

\end{document}